# Drug-target interaction prediction by integrating heterogeneous information with mutual attention network

Yuanyuan Zhang [a], Yingdong Wang[a], Chaoyong Wu[b], Lingmin Zhan[a], Aoyi Wang[a], Caiping Cheng[a], Jinzhong Zhao[a], Wuxia Zhang[a, *], Jianxin Chen[b, *], Peng Li[a, *]

[a] Shanxi key lab for modernization of TCVM, College of Basic Sciences, Shanxi Agricultural University, Taigu 030801, China

[b] School of Traditional Chinese Medicine, Beijing University of Chinese Medicine, Beijing 100029, China

*Corresponding author.

E-mail addresses: wuxia200758@163.com (W, Zhang), cjx@bucm.edu.cn (J. Chen), lip@sxau.edu.cn (P. Li)

**Abstract**

Identification of drug-target interactions is an indispensable part of drug discovery. While conventional shallow machine learning and recent deep learning methods based on chemogenomic properties of drugs and target proteins have pushed this prediction performance improvement to a new level, these methods are still difficult to adapt to novel structures. Alternatively, large-scale biological and pharmacological data provide new ways to accelerate drug-target interaction prediction. Here, we propose DrugMAN, a deep learning model for predicting drug-target interaction by integrating multiplex heterogeneous functional networks with a mutual attention network (MAN). DrugMAN uses a graph attention network-based integration algorithm to learn network-specific low-dimensional features for drugs and target proteins by integrating four drug networks and seven gene/protein networks, respectively. DrugMAN then captures interaction information between drug and target representations by a mutual attention network to improve drug-target prediction. DrugMAN achieves the best prediction performance under four different scenarios, especially in real-world scenarios. DrugMAN spotlights heterogeneous information to mine drug-target interactions and can be a powerful tool for drug discovery and drug repurposing.

## 1. Introduction

Elucidating mechanistic actions of drugs is one of the critical tasks in drug discovery, necessary to identify on-target drugs and new therapeutic targets, avoid unwanted off-target effects, and improve the success rate in clinical trials [1]. Identification of interactions between drugs and their biological targets is the core step of exploring drug mechanisms of action. In these years, both experimental and computational approaches are frequently employed to study drug-target interactions. For instance, direct biochemical assays label the protein or small molecule of interest and directly detect the binding affinity [2]. Molecular dynamics (MD) and docking simulations model potential ligand-target binding configurations with low-energy states [3,4]. Machine learning (ML) and artificial intelligence (AI) models learn molecular representation from chemical structures and capture the complex nonlinear relationships between drugs and targets [5-9]. Till now, there have been many specialist databases providing data on drug-target interactions, such as DrugBank [10], BindingDB [11], ChEMBL [12], and Comparative Toxicogenomics Database (CTD) [13]. These approaches or technologies highlight chemogenomic information to formalize binding interactions between chemicals and targets and neglect other biological information for drugs and protein targets.

Different from chemogenomic-based models, network-based models encode drug and target representations by integrating heterogeneous information from multiplex functional interaction networks, such as inducible gene expression, drug side effects, related diseases, and genetic associations [14]. Intuitively, network-based methods that integrate more information for both drugs and targets could be adept at mining drug-target interactions. In recent years, various network-based methods have been developed based on heterogeneous biological networks to mine drug-target interactions and achieved promising results. For example, DTINet combines Random walk with restart (RWR) and diffusion component analysis (DCA) to learn low-dimensional drug and target representations from heterogeneous networks and predict drug-target interactions using inductive matrix completion [15]. NeoDTI integrates different networks and automatically learns topology-preserving representations of drugs and targets to facilitate drug-target prediction [16].

Despite these promising effects, two challenges remain for network-based methods. (i) An excellent graph embedding method to learn drug and target features. Biological interaction networks from real-world biomedical high-through data inevitably have varied false positives and -negatives while preserving meaningful functional links. A solid graph embedding method should be scalable in both the size and quantity of input networks and learn low-dimensional node features that can reflect the functional and topological properties of all heterogeneous networks. (ii) An embedded module to connect the interaction information between drug and target representations. The network-specific features characterize drug and target proteins in the existing biological big data resources. It should be emphasized to learn the interaction patterns between drugs and targets bridged by these underlying functional media. A simple concatenation of drug and target features is quite inefficient. This problem is similar to the natural language processing (NLP) problem of integrating multiple word embedding into a sentence representation, which has been broken through by the attention mechanism [17].

To address these limitations, we here present a novel deep learning model, termed DrugMAN, which extracts accurate network-specific features of drugs and target proteins by a scalable graph attention network-based integration algorithm and captures interaction patterns between drugs and targets by a mutual attention network to improve drug-target prediction. We have evaluated the performance of DrugMAN and other state-of-the-art methods in real-world applications and found that DrugMAN outperforms both chemogenomics and network-based methods for predicting drug-target interactions under different distributions of test and training datasets.

## 2. Methods

*2.1. Experimental setting*

*2.1.1. Construction of drug-target interaction dataset*

The success of mechanism-based drug discovery depends on the definition of the drug target. To improve the accuracy and reliability of drug discovery, we selected the known drug-target pairs that have been rigorously validated through experiments or supported by extensive literature as the gold-standard data. Drug-target interaction data are collected from five public sources including, Drugbank, map of Molecular Targets of Approved drugs (MTA), CTD, ChEMBL and BindingDB. MTA is a manually curated dataset that contains 1,578 US FDA-approved drugs and 893 human and pathogen-derived biomolecules [18], of which 667 human-genome-derived proteins targeted by 1194 drugs for human diseases are used in the present analysis. We first collect all drug-target pairs from the Drugbank and MTA. Drugs in the form of inorganic salts are removed. Then we collate drug-target pairs in CTD, ChEMBL and BindingDB corresponding to drugs in Drugbank and MTA. All these data are combined and further filtered by the kinetic constants Ki, Kd, IC50 and EC50. Finally, we choose thresholds of $\leq 10^3$ nM to obtain 20,565 drug-target binding data, corresponding to 5135 drugs and 2894 protein targets. All drug compound SMILES and amino acid sequences of targets are obtained from PubChem and UniProt, respectively. The PubChem CID and the gene Entrez ID are used as the unique identifiers for drugs and targets, respectively. When datasets are used for evaluating models in the prediction of drug-target interactions, they are balanced with validated positive interactions and an equal number of negative samples randomly obtained from unseen pairs. The dataset is randomly divided into training, validation and test sets with a 7:1:2 ratio.

*2.1.2. Construction of cold start dataset*

We set up drug cold-start, target cold-start and both cold-start scenarios to evaluate the performance of the model in the real-world drug-target interaction prediction. For cold-start scenarios, each dataset is also randomly divided into training, validation and test sets with a 7:1:2 ratio similar to warm-start.

*2.2. Network acquisition and preprocessing*

*2.2.1 Data acquisition*

We download the latest drug-disease and gene-disease data from the CTD website (https://ctdbase.org/downloads/) [13]. All drug compound SMILES are obtained from PubChem. The latest drug-side effect data are collected from the SIDER database (https://sideeffects.embl.de/) [19]. The gene expression signatures induced by chemical and genetic perturbations are collected from the Cmap Database (https://clue.io/) [20]. The gene-pathway data are downloaded from Reactome, a knowledgebase of biological pathways, reactions, proteins and molecules (https://reactome.org) [21]. The gene-chromosomal locations are curated from Gene, a searchable database of gene-specific contents in the national center for Biotechnology Information (NCBI) (http://www.ncbi.nlm.nih.gov/gene) [22]. The protein-protein interaction data are collected from the STRING network (https://string-db.org) [23]. The gene co-expression data are curated from the Genotype-Tissue Expression (GTEx) dataset (https://www.gtexportal.org) [24]. The sequences of reviewed human proteins are collected from UniProt (https://www.uniprot.org/) [25]. DrugBank is the most widely used drug information resource, which maps drugs to pharmacological targets (https://go.drugbank.com/releases/latest) [10]. MTA is a manually curated dataset that contains 1,578 US FDA-approved drugs and 893 human and pathogen-derived biomolecules [18]. BindingDB dataset is a web-accessible database of compound-target interactions with experimentally validated binding affinities (https://www.bindingdb.org/bind/index.jsp) [11]. ChEMBL is a database of drug small molecules and their biological activity information, including clinical experimental drugs and FDA-approved drugs for therapeutic targets and indications. (https://chembl.gitbook.io/chembl-interface-documentation/downloads) [12].

*2.2 2 Network preprocessing*

Four types of drug networks are applied in this work. The disease-based drug association network is constructed by connecting two drugs related to the same diseases, which are extracted from the Comparative Toxicogenomics Database (CTD) [13]. The side effect-based drug network is built by linking two drugs related to the same side effects from the SIDER database Version 2 [19], The edge weight (i.e. drug similarity between two drugs) in the two networks is calculated by the Jaccard similarity method. The transcriptome-based drug similarity network is constructed by calculating the *Pearson* correlation between gene expression profiles induced by drugs. The transcriptome-based drug network retains edges with similarity scores equal to or above 0.2. The gene expression signatures are collected from the Cmap Database [20]. For each drug, to get a unique signature that accurately measures the drug activity, we combined gene expression signatures of each drug to produce a consensus gene signature by the weighted average algorithm, which calculates a weighted average of the gene expression signatures of each drug, with coefficients given by a pairwise Spearman correlation matrix between the expression profiles of all signatures [26]. The drug structure similarity network is constructed by calculating pair-wise chemical similarity through the Jaccard similarity, based on the Morgan fingerprints with radius 2 implemented in the RDKit package (version: 2020.09.1.0). The drug structure network keeps drug pairs with similarity scores equal or greater than 0.4. To unify all networks for analysis, chemical names in each network are transformed into Pubchem CIDs. All drug networks are mapped to drugs in the curated drug-target interaction dataset and the detailed information of each network is provided in the supplementary table 1.

To incorporate as much gene functional association information as possible, seven classes of gene interaction networks are curated from different biological repositories. The disease-based gene association network is constructed by connecting two genes related to the same diseases based on the CTD dataset. The pathway-based gene network is curated by linking two genes in the same biological pathways, which are downloaded from Reactome, a knowledgebase of biological pathways, reactions, proteins and molecules (https://reactome.org) [22]. The pathway network keeps gene pairs with a similarity magnitude greater than or equal to 0.2 as edges. The chromosomal location-based gene network is built by connecting two genes in the same cytogenetic bands, which are curated from Gene, a searchable database of gene-specific contents in the National Center for Biotechnology Information (NCBI) (http://www.ncbi.nlm.nih.gov/gene). The three gene networks are constructed based on the assumption that two genes associated with the same biological entities should be more functionally related than two genes associated with different biological entities. The similarity between two genes in these networks is quantified by calculating Jaccard similarity scores. Similar to the transcriptome-based drug network, the transcriptome-based gene similarity network is constructed by calculating the *Pearson* correlation between gene expression profiles induced by genetic perturbations. The transcriptome-based gene network contains gene pairs with edge weights equal to or greater than 0.25. In the Cmap Database, we collect all gene expression signatures for three types of genetic perturbations, including shRNA, CRISPR and OE treatments [20]. For each target gene, to get a unique signature that accurately measures the gene activity, we combined gene expression signatures of each target gene to produce a consensus gene signature by the weighted average algorithm [26]. The gene co-expression network is built by calculating the *Pearson* correlation across gene expression profiles in different tissues from the Genotype-Tissue Expression (GTEx) dataset (https://www.gtexportal.org) [24]. The co-expression network includes gene pairs with a *Pearson* correlation magnitude equal to or greater than 0.5. The Search Tool for Recurring Instances of Neighboring Genes (STRING; https://string-db.org) quantitatively integrates different studies and interaction types into a single integrated score for each gene pair based on the total weight of evidence [23]. To obtain networks that are comparable in size to other networks, the STRING network is filtered for only the top 10% of interactions by interaction scores. The protein sequence similarity network is obtained by calculating pairwise Smith–Waterman scores [27]. The sequences of reviewed human proteins are collected from UniProt (https://www.uniprot.org/) [25]. The sequence similarity network retains edges with pairwise similarity scores greater than or equal to 0.23. Gene names in each network are mapped to the human Entrez gene ID. All gene networks are mapped to genes in the five datasets including Drugbank, MTA, CTD, ChEMBL and BindingDB. Detailed information on each network is provided in Supplementary Table 1.

*2.3. DrugMAN architecture*

*2.3.1. Extract drug and target representations from heterogeneous networks*

We adopt BIONIC (Biological Network Integration using Convolutions), a scalable deep learning framework for network integration to learn the accurate and comprehensive representations of drugs and protein targets from different types of drug and target networks, respectively [18]. Each network is represented by its adjacency matrix $A$ where $A_{ij} = A_{ij} > 0$ if node i and node j share an edge and $A_{ij} = A_{ij} = 0$ otherwise. BIONIC encodes each input network using three graph attention networks (GAT) to sequentially extract the three-order neighbors of each node. Each GAT encoder has 10 heads with a hidden dimension of 68 per head. The GAT block formulation is then given by:

$$H_d^{l+1} = \sigma\left(GAT\left(A, W_g^{(l)}, b_g^{(l)}, H_d^{(l)}\right)\right) \quad (1)$$

Where $W_g^{(l)}$ and $b_g^{(l)}$ are the layer-specific learnable weight matrix and bias vector of GAT, $A$ is the adjacency matrix for network $g$, and $H_d^l$ is the $l$th hidden node representation with $H_d^0 = H_P$, $H_p$ is the initial node feature of one-hot encoded so that each node is uniquely identified. $\sigma$ is a nonlinear function (here is LeakyReLU). The final network-specific features learned by the GAT block can retain both local and global features of the network. The network-specific node features for each network are combined through a weighted and stochastically masked summation to produce combined node features $H_{combined}$. Then, BIONIC maps $H_{combined}$ to a low-dimensional space $F$ through a learned linear transformation. In $F$, each row corresponds to a node with learned features of 512 dimensions. BIONIC reconstructs the network $\hat{A} = F \cdot F^T$ and minimize the discrepancy between the reconstruction and input networks to obtain a high-quality $F$. Finally, we use the BIONIC framework to learn drug and target representations $F_d$ and $F_t$ from four types of drug networks and seven gene/protein networks, respectively.

*2.3.2. Mutual attention network*

We further use the self-attention framework to capture pairwise interactions between drug and target features. As input to the mutual attention network, the network-specific drug and target representations $F_d$ and $F_t$ are combined into a new matrix $F_{dt}$:

$$F_{dt} = concat(F_d, F_t, \text{axis} = 0) \qquad (2)$$

$F_{dt}$ has a shape of $2 \times 512$, where the first and second rows are the drug and target representation with 512-dimensional features, respectively. $F_{dt}$ is fed into sequential transformer encoder units [17] to effectively learn the interrelated information between drug and target features:

$$F_{dt}^{l+1} = \sigma(atten(F_{dt}^l, W^l, b^l)) \qquad (3)$$

Where each layer $l$ corresponds to a transformer encoder unit and consists of a self-attention layer and feed-forward neural network layer. W and b are learnable weight matrices and bias vectors in the $l$th transformer encoder unit. $F_{dt}^l$ is the $l$th hidden feature matrix and $F_{dt}^0$ is the initial matrix from the network encoder. σ is the activation function ReLU.

In $F_{dt}$, the updated drug and target representations $F_d$ and $F_t$ are directly concatenated into the joint drug-target pair representation $F_{pair}$ with a dimension of 1024. $F_{pair}$ is then inputted into a classification layer which is a fully connected linear layer connected by a sigmoid output function:

$$F_{pair}^{l+1} = \sigma(W^l F_{pair}^l + b^l)) \qquad (4)$$

$$p = \text{sigmoid}(W_o F_{pair} + b_o) \qquad (5)$$

$W^l$ and $b^l$ are learnable weight matrix and bias vector in linear layers. $W_o$ and $b_o$ are learnable weight matrix and bias vector in the sigmoid layer. $p$ represents the drug-target interaction probability. The binary classification is optimized by minimizing the cross-entropy function as follows:

$$loss = -\sum_i (y_i log(p_i) + (1-y_i)log(1-p_i)) \qquad (6)$$

Where $y_i$ is the ground-truth label of the ith drug-target pair, $p_i$ is its output probability by the model.

*2.4. Implementation*

DrugMAN is implemented in Python 3.8 and PyTorch2.0.0 [29], along with functions from Scikit-learn 1.3.0 [30], Numpy 1.25.2 [31], and Pandas 2.0.3 [32]. The batch size is set to 512 and the Adam optimizer is used with a learning rate of 3e-5. We allow the model to run for at most 400 epochs for all datasets and adjust the learning rate with a cosine annealing strategy [33] before 20 epochs. The best performing model is selected at the epoch giving the best AUROC score on the validation set, which is then used to evaluate the final performance on the test set. In the mutual attention network, the attention block contains five sequential Transformer encoders with eight heads in each self-attention layer. The drug-target pair representation is fed into a multi-layer perceptron consisting of three fully connected linear hidden layers with dimensions [512, 256, 256]. All the hyperparameters mentioned above are carefully manually adjusted to make the model perform optimally. The configuration details analysis are provided in the Supplementary table 2.

*2.5. Baseline*

We compare the performance of DrugMAN with that of the five baseline models including SVM, RF, DeepPurpose, DTINet and NeoDTI. Two shallow machine learning methods, SVM and RF, are applied using the drug-target pair representation concatenated by the drug ECFP4 fingerprint and target protein AAC features. DeepPurpose models drug-target interaction using CNN to encode drug molecular graphs and protein sequences. The learned drug and protein representation vectors are combined with a simple concatenation and processed by a binary classification layer. DTINet is a network-based model by combining restarted random walk and diffusion component analysis to learn low-dimensional feature representations of drugs and targets from heterogeneous networks and predicts drug-target interaction using inductive matrix completion. NeoDTI is an end-to-end network-based model to integrates diverse heterogeneous networks and automatically learns topology-preserving representations of drugs and targets to predict drug-target interactions.

*2.6. Code availability*

The source code and implementation details of DrugBAN are freely available at https://github.com/lipi12q/DrugMAN

**3. Result**

*3.1. DrugMAN architecture*

DrugMAN contains two main networks. The first network intends to learn accurate and comprehensive representations for drugs and protein targets from heterogeneous drug and gene/protein networks by the network integration algorithm BIONIC (Biological

Network Integration using Convolutions), which outperforms the existing state-of-the-art network embedding methods [18]. The core idea of BIONIC is to characterize the neighborhoods of network nodes through sequential graph attention networks (GAT) to learn network-specific integrated features. The second network captures the relevant information in a drug-target pair and learns the predictive score for drug-target interaction. We treat the drug and target in a drug-target pair as words in a sentence. The drug and target representations are processed by the mutual attention network that utilizes a series of transformer encoders to apprehend interaction information between the drug and target. The updated drug and target features are concatenated to form the drug-target pair representation, which is then run through a sequence of fully connected classification layers to obtain the predictive score, indicating the probability of drug-target interaction.

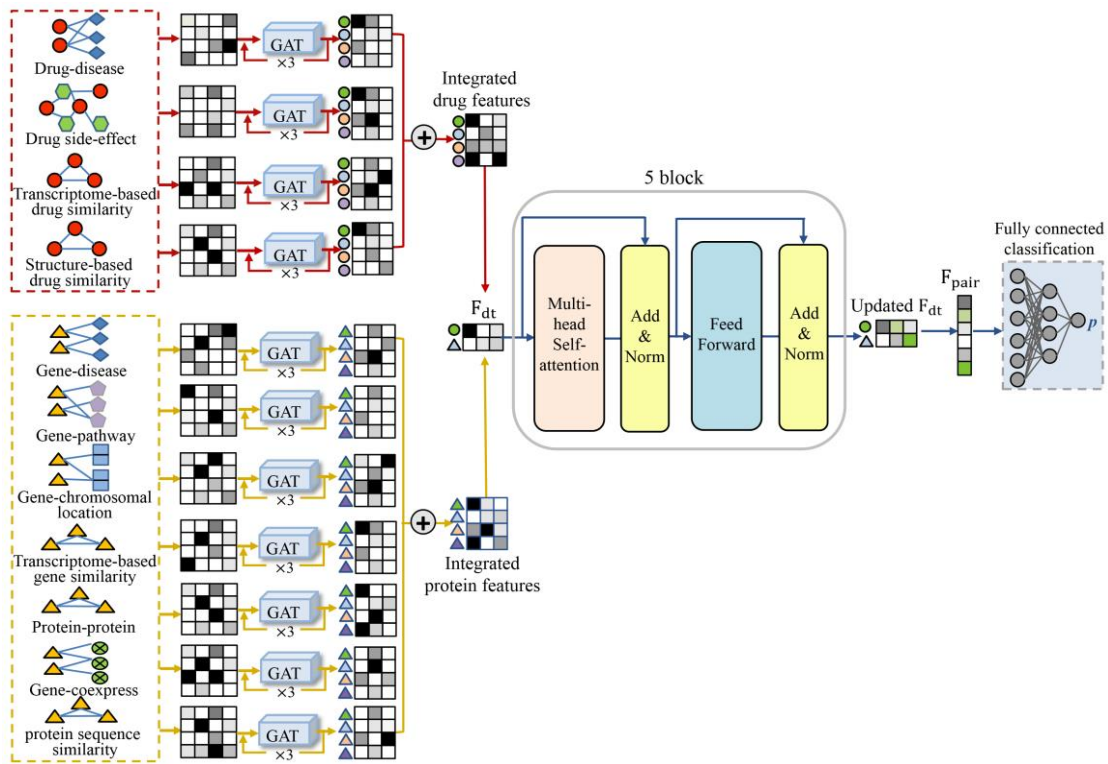

**Figure 1. DrugMAN framework.** DrugMAN contains two main parts. The first part encodes network-specific drug and target features from heterogeneous drug and gene/protein networks through sequential graph attention networks (GAT). The combined drug and target features ($F_{dt}$) are fed into the second part to learn the updated $F_{dt}$ by the five transformer encoders. The updated $F_{dt}$ captures interaction information between the drug and the target. Then the drug and target features in the updated $F_{dt}$ are concatenated to drug-target pair representation ($F_{pair}$), which is input to the fully connected classification layer to calculate the drug-target binding probability score.

*3.2. Evaluation criteria*

We curate drug-target interaction data that meet the rigorous standards from five common sources including DrugBank, map of Molecular Targets of Approved drugs (MTA), CTD, ChEMBL, and BindingDB (see methods for details). To train the model, the dataset is randomly divided into training, validation, and test sets with a 7:1:2 ratio. The training set is used to fit the model, and the test set is used to evaluate the model's performance. It should be noted that in real-world scenarios, drug–target pairs that need to be predicted are often unseen and dissimilar to any pairs in the training data. To evaluate the performance of the model in real-world applications, we set three different scenarios to simulate the real-world prediction for drug-target interactions: (i) scenario of drug cold-start (drug-cold), in which drugs from the test set are absent in the training data, (ii) scenario of target cold-start (target-cold), where targets from the test set are not present in the training data, and (iii) both cold-start (both-cold), where both drugs and targets from the test set are missing in the training data.

The area under the receiver operating characteristic curve (AUROC), the area under the precision-recall curve (AUPRC), and F1-score are used as the major metrics to evaluate the model classification performance. For each experiment, the best-performing model is the one with the best AUROC on the validation set. We conduct five independent runs with different random seeds for each dataset split. The average value of AUROC, AUPR, and F1-score of five runs are used as indicators for model evaluation.

*3.3. Evaluation of DrugMAN and baselines*

We first compare the performance of DrugMAN with three chemoinformatic baseline models, support vector machine (SVM), random forest (RF), and DeepPurpose. As shown in **Figure 2**, DrugMAN consistently outperforms the three chemoinformatic baselines in terms of AUROC, AUPRC, and F1-score. To discern whether the superiority of DrugMAN over these chemoinformatic methods is due to its additional heterogeneous network information, two classical network-based models DTINet and NeoDTI are introduced with the same network data as DrugMAN. We find that DTINet and NeoDTI only perform better than SVM but underperform both RF and DeepPurpose. The results indicate integrating more information cannot necessarily guarantee better performance in predicting drug-target interactions for network-based models compared to chemoinformatic methods [34]. More importantly, the comparison between network-based methods demonstrates that DrugMAN yields superior performance than the two state-of-the-art models. Specifically, it outperforms DTINet and NeoDTI by 16.6% and 7.1% in AUROC, 12.8% and 8.1% in AUPRC, and 18.3% and 8.3% in F1-score, respectively. The results show that DrugMAN better integrates data from different types of networks and captures relevant information in drug-target pairs.

We further evaluate the robustness of DrugMAN in three different real-world scenarios. As expected, compared to the normal random (warm-start) condition, the performance of all models drops significantly due to less information overlap between training and test data in cold-start scenarios. Even so, DrugMAN still achieves the best performance against other state-of-the-art baselines including chemoinformatic and network-based models in drug-cold (AUROC = 0.910, AUPRC = 0.921 and F1-score = 0.835), target-cold (AUROC = 0.922, AUPRC = 0.931 and F1-score = 0.846) and both-cold (AUROC = 0.850, AUPRC = 0.861 and F1-score = 0.776). The results confirm the generalization ability of DrugMAN in the drug-target interaction prediction.

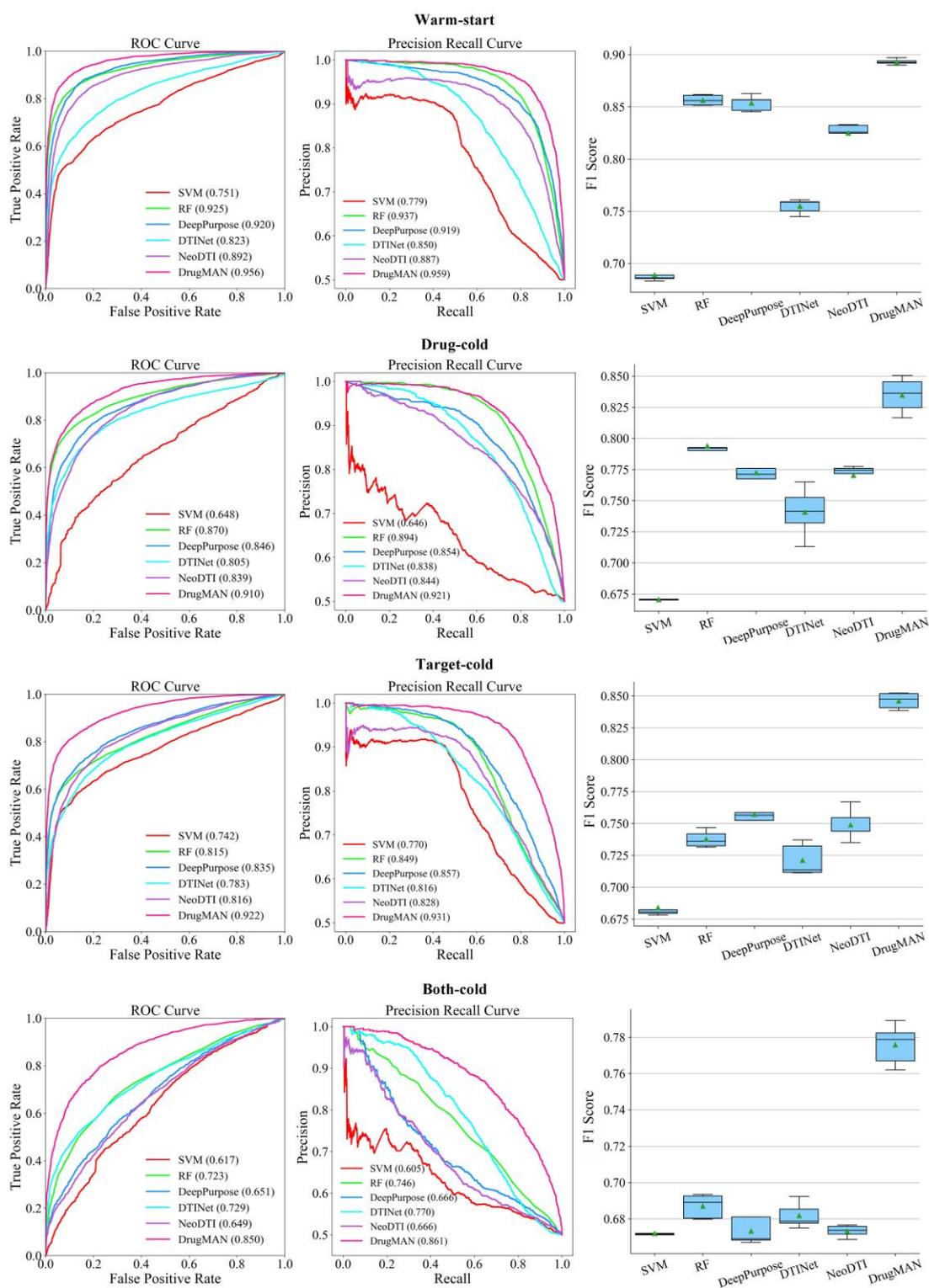

**Figure 2. Comparison of DrugMAN to state-of-the-art methods.** We compare the prediction performance of DrugMAN to that of five baselines, including three

chemoinformatic models SVM, RF and DeepPurpose, and two network-based models DTINet and NeoDTI. The rows from top to bottom correspond to four scenarios: warm-start, drug-cold, target-cold and both cold, respectively. The columns from left to right correspond to three metrics: the receiver operating characteristic curve, precision-recall curve and F1 Score. The box plots show the median as the centre lines and the mean as green triangles for five random runs. The minima and lower percentile represent the worst and second-worst scores. The maxima and upper percentile indicate the best and second-best scores.

*3.4. Evaluation of different network embedding methods*

The network integration algorithm is the essential part of DrugMAN for extracting drug and target features from heterogeneous networks. Although BIONIC has been proven to perform better than other established network integration methods in various benchmarks as a whole, for a fair comparison in drug-target interaction prediction task, we compare BIONIC with two classical network integration approaches, deepNF a deep learning multi-modal autoencoder and multi-node2vec a multi-network extension of the node2vec model by directly replacing BIONIC in the DrugMAN framework. When DrugMAN uses the drug and target representations learned from deepNF and multi-node2vec, we observe an overall drop in performance in terms of AUROC, AUPRC, and F1-score across all scenarios (**Figure 3**). Especially, we observe that the advantage of BIONIC is even more remarkable in cold-start conditions compared to the random split. For example, in the random testing, with the assistance of BIONIC,

DrugMAN outperforms models with substitution of multi-node2vec by 2.8%, 2.6% and 3.8% in AUROC, AUPRC and F1-score, respectively. In the both-cold testing, this discrepancy has increased to 15.3%, 15.4% and 10.4% in AUROC, AUPRC and F1-score, respectively. The results indicate that BIONIC captures more sophisticated functional and topological information from drug and target heterogeneous networks to power the drug-target interaction prediction compared to the established network integration methods.

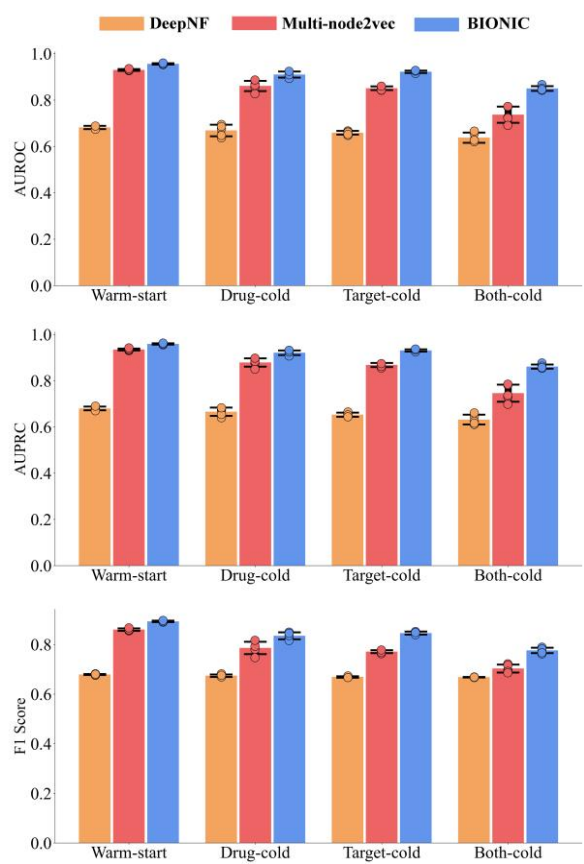

**Figure 3. Evaluation of different network embedding methods.** In DrugMAN, the drug and target embedding module BIONIC is replaced by two other network

integration methods: DeepNF and Multi-node2vec. The vertical bars represent the mean value of five random runs, and the black lines are error bars indicating the standard deviation. The dots indicate performance scores in each random run.

*3.5. Evaluation of integrating and single networks*

To evaluate the impact of each single network on the model predicting ability, for each run we use one single network to produce the drug or target features and fix all other settings in DrugMAN. We find that all single network-based models (including four drug networks and seven protein target networks) show poorer performance compared to the primary DrugMAN (**Supplementary Tables 3 and 4**), confirming the significance of integrating heterogeneous networks for drug-target interaction prediction. For input drug networks, the best performance is observed for the drug structure similarity network-based model with AUROC of 0.949, 0.896, 0.913 and 0.834, AUPRC of 0.953, 0.910, 0.923 and 0.844 and F1-score of 0.883, 0.823, 0.837 and 0.764, in random, drug-cold, target-cold and both-cold testing, respectively (**Supplementary Table 3**). This is reasonable as the structure information is the basis for drug binding to the target. Consistently, in all single protein network-based models, the protein sequence similarity network-based model performs better than or as well as other single network models across all scenarios (**Supplementary Table 4**). Based on the prominent contribution of structural information to DrugMAN, we further examine the performance of the model with only the drug structure similarity network and the protein sequence similarity network as input (DrugMAN$_{STR}$). Encouragingly,

DrugMAN$_{STR}$ outperforms those state-of-the-art structural models including SVM, RF and DeepPurpose, indicating the DrugMAN framework can capture more information from the structural similarity data related to drug-target interactions compared to the established methods. Moreover, compared to DrugMAN$_{STR}$, we can observe that DrugMAN has significant performance improvements with the introduction of non-structural heterogeneous information in all scenarios (**Figure 4**). The discrepancy between DrugMAN$_{STR}$ and the primary DrugMAN can to some extent reflect the contribution of non-structural information to the performance. For example, in the both-cold scenario, DrugMAN outperforms DrugMAN$_{STR}$ by 7.6%, 6.2% and 6.7% in AUROC, AUPRC and F1-score, respectively (**Figure 4**). These results demonstrate the strength of DrugMAN in generalizing prediction performance across different conditions by integrating various pharmacological and biological information.

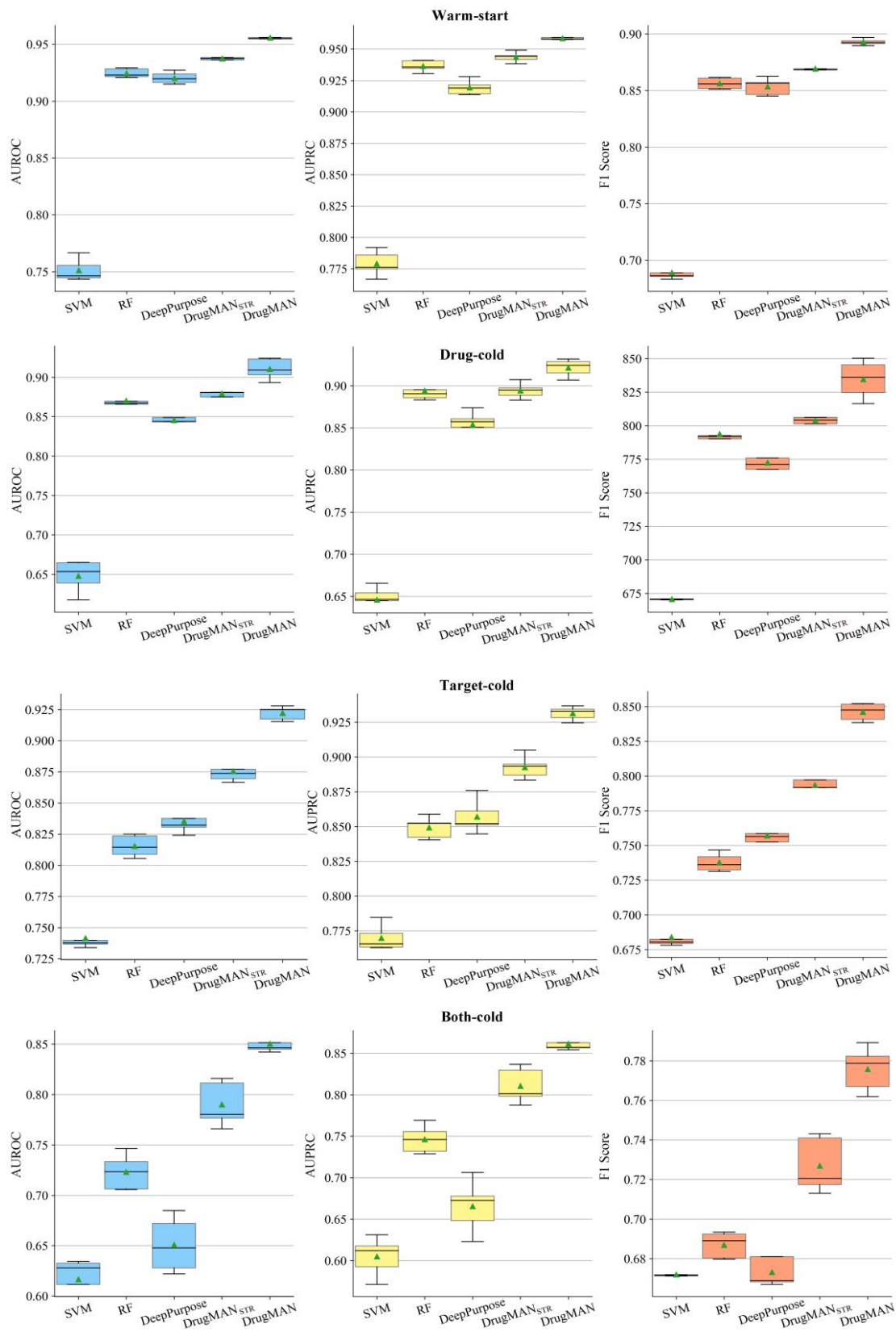

**Figure 4. Evaluation of DrugMAN based on the drug structure similarity network**

**and the protein sequence similarity network.** DrugMAN with only the drug structure similarity network and the protein sequence similarity network as input (DrugMAN$_{STR}$) outperforms three state-of-the-art chemoinformatic models but underperforms DrugMAN. The box plots show the median as the center lines and the mean as green triangles. The minima and lower percentile represent the worst and second-worst scores. The maxima and upper percentile indicate the best and second-best scores.

*3.6. Ablation study*

We perform an ablation study to investigate the impact of the mutual attention network on DrugMAN. As shown in **Figure 5**, the introduction of the mutual attention network has significantly improved the performance of DrugMAN in all scenarios. The results indicate that the mutual attention network can capture the pairwise interaction information for drug-target interaction prediction.

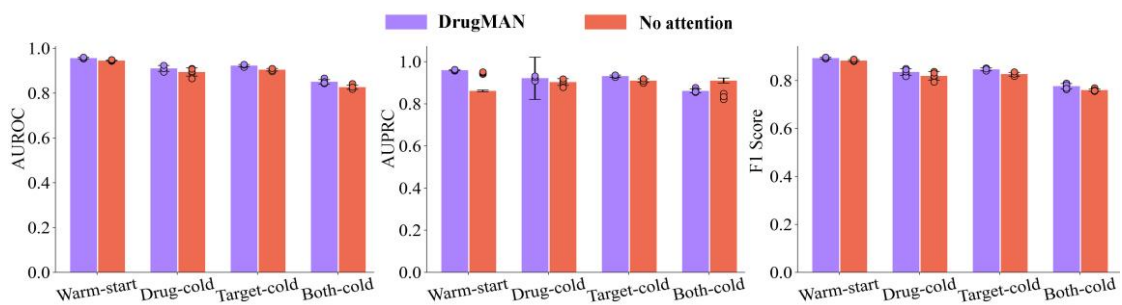

**Figure 5. Ablation study for DrugMAN without attention mechanisms**. Performance comparison of DrugMAN with and without attention mechanisms (No attention) in different scenarios. The vertical bars represent the mean of five random

runs, and the black lines are error bars indicating the standard deviation. The dots indicate performance scores in each random run.

## 4. Discussion

In this work, we develop a new framework, DrugMAN, to integrate drug and protein target information from multiplex biological networks to mine drug-target interactions. DrugMAN achieves superior performance over both state-of-the-art chemogenomics and network-based models. Especially, the robustness of DrugMAN has been confirmed in different real-world scenarios. DrugMAN's effectiveness can be attributed to two intrinsic advantages: the sophisticated network embedding module for learning suitable drug and target features and the mutual attention block to capture interaction information between drugs and targets.

The main challenge of network embedding is how to encode accurate node features from heterogeneous networks with high-dimensional, incomplete and noisy traits. DrugMAN takes BIONIC, the latest deep learning-based network integration algorithm, which first characterizes the topology of each individual network by applying a GAT algorithm, and then formalizes a low-dimensional representation by combining the features learned from each individual network to approximate the initial networks. The network-specific integrated features learned by BIONIC can reflect both functional and topological properties of heterogeneous networks and excel other unsupervised methods in a range of downstream tasks [18]. We demonstrate that DrugMAN can achieve substantial improvement over the state-of-the-art network-based methods for drug-target interaction prediction (**Figure 2**). Consistently, we here compare BIONIC with two classical network integration approaches, deepNF and multi-node2vec in the DrugMAN framework (**Figure 3**), indicating that BIONIC is more adaptive for drug-target interaction prediction compared to the established network integration methods. In addition, we find when DrugMAN uses single drug or target networks as input, the predicting performance is greatly inferior compared to DrugMAN with multiple networks (**Supplementary Tables 3 and 4**), indicating BIONIC can produce accurate integrated drug and target features from heterogeneous networks for drug-target interaction prediction.

Most existing drug-target prediction models learn drug and target representations using their separate encoders and ignore mutual impacts between the targets and drugs [35,36]. The pairwise interaction information between drugs and targets is explicitly important for drug-target interaction prediction [37,38]. We use a mutual attention network that utilizes a series of transformer encoders to capture interaction patterns between drugs and targets. We demonstrate that the introduction of the mutual attention block in the DrugMAN architecture significantly improves the performance of DrugMAN in all scenarios (**Figure 5**). In summary, DrugMAN can provide a powerful and useful tool to facilitate drug discovery and drug repositioning. We will further extend DrugMAN by integrating more biological networks and mine meaningful drug-target pairs through wet lab experiments.

**CRediT authorship contribution statement**

Yuanyuan Zhang: Conceptualization, Methodology, Software, Formal analysis, Visualization, Data curation, Writing-original draft. Chaoyong Wu: Conceptualization, Methodology, Software, Formal analysis, Visualization, Data curation, Writing-original draft. Linmin Zhan: Conceptualization, Methodology, Software, Validation, Data curation. Yingdong Wang: Investigation, Resources, Data curation, Formal analysis, Validation. Aoyi Wang: Investigation, Resources, Data curation, Formal analysis, Visualization. Caiping Cheng: Conceptualization, Investigation, Supervision, Validation. Jinzhong Zhao: Conceptualization, Investigation, Supervision, Validation. Wuxia Zhang: Conceptualization, Resources, Supervision, Validation, Writing – review & editing. Jianxin Chen: Conceptualization, Methodology, Writing – review & editing. Peng Li: Conceptualization, Methodology, Formal analysis, Investigation, Supervision, Project administration, Writing-original draft, Writing – review & editing.

**Conflict of interest**

The authors declare that they have no known competing financial interests or personal relationships that could have appeared to influence the work reported in this paper.

**Funding**

This research was supported by the National Natural Science Fund of China (No. 82274363), the Fundamental Research Program of Shanxi Province (No. 20210302124129), and the Distinguished and Excellent Young Scholars Cultivation Project of Shanxi Agricultural University (No. 2022YQPYGC09).

# References

[1] Hughes J.P., Rees S., Kalindjian S.B., et al., Principles of early drug discovery. Br J Pharmacol. 162 (2011) 1239-1249. https://doi.org/10.1111/j.1476-5381.2010.01127.x.

[2] Schenone M., Dančík V., Wagner B.K., et al., Target identification and mechanism of action in chemical biology and drug discovery. Nat Chem Biol. 9 (2013) 232-240. http://dx.doi.org/10.1038/nchembio.1199.

[3] De Vivo M., Masetti M., Bottegoni G., et al., Role of Molecular Dynamics and Related Methods in Drug Discovery. J Med Chem. 59 (2016) 4035-4061. http://dx.doi.org/10.1021/acs.jmedchem.5b01684.

[4] Meng X.Y., Zhang H.X., Mezei M., et al.., Molecular docking: a powerful approach for structure-based drug discovery. Curr Comput Aided Drug Des. 7 (2011) 146-157. http://dx.doi.org/10.2174/157340911795677602.

[5] Chen H., Engkvist O., Wang Y., et al., The rise of deep learning in drug discovery. Drug Discov Today. 23 (2018) 1241-1250. http://dx.doi.org/10.1016/j.drudis.2018.01.039.

[6] Huang K., Fu T., Glass L.M., et al., DeepPurpose: a deep learning library for drug-target interaction prediction. Bioinformatics. 36 (2021) 5545-5547. http://dx.doi.org/10.1093/bioinformatics/btaa1005.

[7] Chatterjee A., Walters R., Shafi Z., et al., Improving the generalizability of protein-ligand binding predictions with AI-Bind. Nat Commun. 14 (2023) 1989. http://dx.doi.org/10.1038/s41467-023-37572-z.

[8] Wang X., Cheng Y., Yang Y., et al., Multitask joint strategies of self-supervised

representation learning on biomedical networks for drug discovery. Nature Machine Intelligence. 5 (2023) 445-456. http://dx.doi.org/10.1038/s42256-023-00640-6.

[9] Zeng X., Xiang H., Yu L., et al., Accurate prediction of molecular properties and drug targets using a self-supervised image representation learning framework. Nature Machine Intelligence. 4 (2022) 1004-1016. http://dx.doi.org/10.1038/s42256-022-00557-6.

[10] Wishart D.S., Knox C., Guo A.C., et al., DrugBank: a knowledgebase for drugs, drug actions and drug targets. Nucleic Acids Res. 36 (2008) D901-906. http://dx.doi.org/10.1093/nar/gkm958.

[11] Gilson M.K., Liu T., Baitaluk M., et al., BindingDB in 2015: A public database for medicinal chemistry, computational chemistry and systems pharmacology. Nucleic Acids Res. 44 (2016) D1045-1053. http://dx.doi.org/10.1093/nar/gkv1072.

[12] Gaulton A., Hersey A., Nowotka M., et al., The ChEMBL database in 2017. Nucleic Acids Res. 45 (2017) D945-d954. http://dx.doi.org/10.1093/nar/gkw1074.

[13] Davis A. P., Grondin C. J., Johnson R. J., et al., Comparative Toxicogenomics Database (CTD): update 2021. Nucleic Acids Res. 49 (2021) D1138-d1143. http://dx.doi.org/10.1093/nar/gkaa891.

[14] Zong N., Wong R.S.N., Yu Y., et al., Drug-target prediction utilizing heterogeneous bio-linked network embeddings. Brief Bioinform. 22 (2021) 568-580. http://dx.doi.org/10.1093/bib/bbz147.

[15] Luo Y., Zhao X., Zhou J., et al., A network integration approach for drug-target interaction prediction and computational drug repositioning from heterogeneous information. Nat

Commun. 8 (2017) 573. http://dx.doi.org/10.1038/s41467-017-00680-8.

[16] Wan F., Hong L., Xiao A., et al., NeoDTI: neural integration of neighbor information from a heterogeneous network for discovering new drug-target interactions. Bioinformatics. 35 (2019) 104-111. http://dx.doi.org/10.1093/bioinformatics/bty543.

[17] Vaswani A., Shazeer N., Parmar N., et al., Attention is all you need, Proceedings of the 31st International Conference on Neural Information Processing Systems, Curran Associates Inc., Long Beach, California, USA, 2017, pp. 6000–6010.

[18] Santos R., Ursu O., Gaulton A., et al., A comprehensive map of molecular drug targets. Nat Rev Drug Discov. 16 (2017) 19-34. http://dx.doi.org/10.1038/nrd.2016.230.

[19] Kuhn M., Letunic I., The SIDER database of drugs and side effects. Nucleic Acids Res. 44 (2016) D1075-1079. http://dx.doi.org/10.1093/nar/gkv1075.

[20] Subramanian A., Narayan R., Corsello S. M., et al., A Next Generation Connectivity Map: L1000 Platform and the First 1,000,000 Profiles. Cell. 171 (2017) 1437-1452.e1417. http://dx.doi.org/10.1016/j.cell.2017.10.049.

[21] Gillespie M., Jassal B., Stephan R., et al., The reactome pathway knowledgebase 2022. Nucleic Acids Res. 50 (2022) D687-d692. http://dx.doi.org/10.1093/nar/gkab1028.

[22] Sayers E. W., Bolton E. E., Brister J. R., et al., Database resources of the National Center for Biotechnology Information in 2023. Nucleic Acids Res. 51 (2023) D29-d38. http://dx.doi.org/10.1093/nar/gkac1032.

[23] Szklarczyk D., Gable A.L., Nastou K.C., et al., The STRING database in 2021: customizable protein-protein networks, and functional characterization of user-uploaded

gene/measurement sets. Nucleic Acids Res. 49 (2021) D605-d612. http://dx.doi.org/10.1093/nar/gkaa1074.

[24] GTEx Consortium, The Genotype-Tissue Expression (GTEx) project. Nat Genet. 45 (2013) :580-585. https://doi:10.1038/ng.2653.

[25] Uiport. Consortium, UniProt: the universal protein knowledgebase in 2021. Nucleic Acids Res. 49 (2021) D480-d489. http: //dx.doi.org/10.1093/nar/gkaa1100.

[26] Smith I., Greenside P.G., Natoli T., et al., Evaluation of RNAi and CRISPR technologies by large-scale gene expression profiling in the Connectivity Map. PLoS Biol. 15 (2017) e2003213. http://dx.doi.org/10.1371/journal.pbio.2003213.

[27] Smith T.F., Waterman M.S., Identification of common molecular subsequences. J Mol Biol. 147 (1981) 195-197. http: //dx.doi.org/10.1016/0022-2836(81)90087-5.

[28] Forster D.T., Li S.C., Yashiroda Y., et al., BIONIC: biological network integration using convolutions. Nat Methods. 19 (2022) 1250-1261. http://dx.doi.org/10.1038/s41592-022-01616-x.

[29] Paszke A., Gross S., Massa F., et al., PyTorch: An Imperative Style, High-Performance Deep Learning Library, Neural Information Processing Systems, 2019.

[30] Pedregosa F., Varoquaux G., Gramfort A., et al., Scikit-learn: Machine Learning in Python. J. Mach. Learn. Res. 12 (2011) 2825–2830.

[31] Harris C.R., Millman K.J., Array programming with NumPy. Nature. 585 (2020) 357-362. http://dx.doi.org/10.1038/s41586-020-2649-2.

[32] The pandas development team. pandas-dev/pandas: Pandas 2.0.3. Zenodo

https://doi.org/10.5281/zenodo.8092754 (2023)

[33] Loshchilov I., Hutter F., SGDR: Stochastic Gradient Descent with Warm Restarts. ICLR. (2016) http://dx.doi.org/DOI:10.48550/arXiv.1608.03983.

[34] Zong N., Li N., Wen A., et al., BETA: a comprehensive benchmark for computational drug-target prediction. Brief Bioinform. 23 (2022) . http: //dx.doi.org/10.1093/bib/bbac199.

[35] Sydow D., Burggraaff L., Szengel A., et al., Advances and Challenges in Computational Target Prediction. Nature. 59 (2019) 1728-1742. http://dx.doi.org/10.1021/acs.jcim.8b00832.

[36] Chen X., Yan C.C., Zhang X., et al., Drug-target interaction prediction: databases, web servers and computational models. Brief Bioinform. 17 (2016) 696-712. http://dx.doi.org/10.1093/bib/bbv066.

[37] Bai P., Miljković F., John B., et al., Interpretable bilinear attention network with domain adaptation improves drug–target prediction. Nature Machine Intelligence. 5 (2023) 126-136. http://dx.doi.org/10.1038/s42256-022-00605-1.

[38] Li F., Zhang Z., Guan J., et al., Effective drug-target interaction prediction with mutual interaction neural network. Bioinformatics. 38 (2022) 3582-3589. http://dx.doi.org/10.1093/bioinformatics/btac377.